\documentstyle[12pt]{article}
\begin{document}

\baselineskip=24pt plus 2pt
\hfill{\sl NCKU-HEP/95-08}

\hfill{\sl Preliminary version}
\vspace{5mm}
\begin{center}
{\large \bf  The Post-Newtonian Limit of Dilaton Gravity} \\
\vspace{5mm}
Rue-Ron Hsu$^{\dagger}$\footnote{E-mail address: rrhsu@mail.ncku.edu.tw}, Bokai
Yang$^{\dagger}$ and Chin-Rong Lee$^{\ddagger}$\footnote{E-mail address:
phycrl@ccunix.ccu.edu.tw}\\
$^{\dagger}$Department of Physics, National Cheng Kung University, Tainan, \\
{}~~Taiwan 701, Republic of China \\
$^{\ddagger}$Department of Physics, National Chung Cheng University, Chia-Yi,
\\
{}~~Taiwan 62117, Republic of China \\
\end{center}
\vspace{5mm}

\begin{center}
{\bf ABSTRACT}
\end{center}
We study the post-Newtonian limit of a generalized dilaton gravity in which
gravity is coupled to dilaton and eletromagnetic fields.  The field equations
are derived using the post-Newtonian scheme, and the approximate solution is
presented for a point mass with electric and dilaton charges. The result
indicates that the dilaton effect can be detected, in post-Newtonian level,
using a charged test particle but not a neutral one.  We have also checked that
the approximate solution is indeed consistent with the weak field expansion of
charged dilaton black hole solution in the harmonic coordinate.
\newpage
\section{Introduction}

It is of interest to investigate how the properties of black holes are modified
when the low-energy effective actions of superstring theories~\cite{1}-\cite{5}
are considered.  Some new black hole solutions have been obtained in the
low-energy string theories in which the Kalb-Ramond field, dilaton field and
gauge field are incorporated with gravity~\cite{6}-\cite{32}.   Above all, when
non-minimally coupled dilaton and $U(1)$ gauge field are included, the
effective theory is called dilaton gravity.  The dilaton gravity theory has
been studied extensively in recent years due to many interesting properties in
this theory.
The non-minimal coupling allows to violent the no-hair theorem which offers the
uniqueness of the black hole solutions of the Einstein theory.  In the extremal
limit, the black hole is on the edge of becoming naked singularity and should
be regarded as elementary particles.~\cite{25}\cite{26}.  Moreover, a very
useful property of extremal black hole solution is that the metric split into a
direct product of 2-d solution with a 2-sphere of constant radius.  That
naturally leads to a reduced 2-dimensional model, called CGHS model~\cite{27}.
Studying  a 2-d dilaton gravity instead of a 4-d theory, makes the problem more
tractable.  In fact, many works have devoted to studying the evaporation and
information puzzle of the 4-dimensional black hole, by investigating CGHS
mode~\cite{28}.

However, gravitational theory is a experimental tested theory, such as general
relativity.
Recently, Damour and Polyakov~\cite{33} studied the violation of the
equivalence principle in a string inspirited model-dilaton gravity. They found
that the violation is well below the present experimental limit.  It means that
the high precision tests of the equivalence principle can be viewed as
low-energy window on string-scale physics.  That provides a new motivation for
improving the experimental tests of Einstein's Equivalence Principle.

It is interested to ask whether the dilaton gravity will imply some new
observable effects in the post-Newtonian limit.~\cite{34} These studies will
improve our understanding in effective string theories. In this report, we will
derive the field equation of a generalized dilaton gravity, in which gravity,
electromagnetic field and dilaton field are coupled to their source terms
individually in the post-Newtonian limit . We also demonstrate an approximate
solution for a point mass with electric charge and dilaton charge, and compare
it with the weak field expansion of a static, charged dilatonic black hole in
the harmonic coordinate.

The plan of this paper is as follows. In Sec. 2, we review the black hole
solution in the dilaton gravity.  In Sec. 3, field equations of the modified
dilaton gravity are developed in the post-Newtonian scheme, and an approximate
solution of a point mass with both electric and dilaton charges was
demonstrated.  We rewrite the electric dilatonic black hole in the harmonic
coordinate and compare it with the approximate solution in Sec. 4. Finally, we
present some concluding remarks.

\vspace{10mm}
\section{Charged dilatonic black hole }

The dilaton gravity is described by the action,
\begin{equation}
   I = \int d^4x \sqrt{-g} \left[  R - 2(\nabla \phi)^2
                                      - e^{-2\phi} F^2 \right]~~.
\end{equation}
Gibbons and Maede\cite{13} obtained the dyonic black hole solution for the
theory in terms of the non-standard metric form. One of us\cite{30} found the
same solution in terms of the standard metric form,
\begin{equation}
   ds^2 = -\Delta^2 dt^2 + \frac{\sigma^2}{\Delta^2} dr^2
          + r^2 \left( {d\theta}^2
          + {\sin}^2 \theta {d\varphi}^2 \right)~~,
\end{equation}
where
\begin{equation}
   \Delta^2 = 1 - \frac{2 M}{r^2} \sqrt{r^2 + \rho^2}
                + \frac{\beta}{r^2}~~,
\end{equation}
\begin{equation}
   \sigma^2 = \frac{r^2}{r^2 + \rho^2}~~,
\end{equation}
and
\begin{equation}
   \rho = \frac{1}{2 M} ({Q_e}^2 e^{2\phi_0}
          - {Q_m}^2 e^{-2\phi_0})~~,
\end{equation}
\begin{equation}
   \beta = ({Q_e}^2 e^{2\phi_0} + {Q_m}^2 e^{-2\phi_0})~~.
\end{equation}
The electric and magnetic fields are
\begin{equation}
   F_{01} = \frac{Q_e}{r^2} e^{2\phi}~~,
\end{equation}
\begin{equation}
   F_{23} = \frac{Q_m}{r^2}~~.
\end{equation}
and the dilaton field obeys
\begin{equation}
   e^{2\phi} = e^{2\phi_0}(1-\frac{2 \rho}{\sqrt{r^2 + \rho^2}
               + \rho})~~.
\end{equation}
The solution is characterized by mass $M$, electric charge $Q_e$, magnetic
charge $Q_m$ and asymptotic value of the dilaton $\phi_0$.  It is obvious that
the dual transformation can be represented by changing $(Q_e,Q_m,\phi)$ to
$(Q_m,Q_e,-\phi)$ in the solution.

In this form , the properties of black hole are much easier to interpret and to
be compared with those of Reissner-Nordstr\"om black holes.  The structure and
thermodynamical properties of dyonic dilaton black hole are similar to those of
the conventional charged black hole, except for the pure electric or the pure
magnetic cases.  For the extremal charged black hole,
\begin{equation}
M = \frac{1}{\sqrt{2}} (|Q_e| e^{\phi_0} + |Q_m| e^{-\phi_0}),
\end{equation}
the thermodynamical description is inappropriate\cite{15}\cite{29}\cite{30}.
When we set $Q_m=0$ or $Q_e=0$, the solution is reduced to the pure electric or
pure magnetic solutions.  Garfinkle {\it et al.}~\cite{14} also found these
solutions in terms of non-standard metric form .  The description of the pure
electric or pure magnetic black holes as thermal objects break down as the
extreme limit is approached.  Preskill {\it et al.}~\cite{25} and Holzhey {\it
et al.}~\cite{26} suggest that these extreme charged solutions should be
regarded as elementary particles.

Although, the exact spherically symmetric solutions  are found in the massless
dilaton gravity. The exact axially symmetric solution is not found, only the
approximate solution was presented~\cite{21}.  And they only found a numerical
solution for the massive dilaton black hole~\cite{19}\cite{20}.  Therefore,
beside to understand the experimental tested effect in dilaton gravity, the
post-Newtonian scheme also offers a systematic way to find all kinds of
approximate solutions.

\vspace{10mm}
\section{The post-Newtonian scheme}

Recently, the post-Newtonian limit of the superstring effective action was
discussed by Kaehagias~\cite{35}. He studied the post-Newtonian limit and
gravitational radiation of the effective action in which gravity is coupled to
dilaton field and antisymmetric tensor field called axion.  He found that the
theory is identical to Einstein gravity in the post-Newtonian approximation.
Also, he predicted all possible types of radiation in the weak field
approximation.  There exist monopole, dipole, quadrupole, etc. contributions in
the radiation luminosity.

Here, we consider a generalized action of dilaton gravity in which gravity is
coupled to dilaton, electromagnetic field and source terms.
\begin{equation}
I = \int d^4x \sqrt{-g} \biggl\{ \left[ { R\over{2\kappa^2}}
- 2(\nabla \phi)^2- b e^{-2a\phi} F^2 \right]
+{ \cal{L}}_{matter}( g_{\mu\nu},A_{\mu},\phi ) \biggr\} ~,
\end{equation}
with constant parameters $\kappa^2=8\pi G, a=\sqrt{2}\kappa$ and
$b={{\alpha}'\over{16\kappa^2}}$.  $G$ is the gravitational constant and
${\alpha}'$ is the fundamental constant in string theory.  When
$b={1\over8}$~\cite{3}, the action is the effective theory of four-dimensional
heterotic string. When $a=0$, the action reduces to the usual Einstein-Maxwell
scalar theory. The dilaton gravity is the special case when $2\kappa^2=1$,$b=1$
and $a=1$. In this paper, we will use the geomtrized unit, that is $G=1$ and
$c=1$, and we adapt the convention of MTW\cite{36}.

The field equations of the effective string action, eq.(11), are
\begin{equation}
   R_{\mu \nu} = 2\kappa^2\lbrack 2\nabla_{\mu}\phi\nabla_{\nu}\phi +
2be^{-2a\phi}(F_{\mu \rho} F_{\nu}^{~\rho} - \frac{1}{4}g_{\mu \nu}F^2)\rbrack
+\kappa^2(T_{\mu\nu}^{m}-\frac{1}{2}g_{\mu\nu}T^{m})~~,
\end{equation}
\begin{equation}
\nabla_{\mu} (e^{-2a\phi} F^{\mu \nu})~~=-4\pi J^{\nu},
\end{equation}
and
\begin{equation}
   \nabla^2 \phi + \frac{ab}{2} e^{-2a\phi} F^2 = -4\pi\Sigma~~,
\end{equation}
where
$T^{\mu\nu}_{m}=\frac{2}{\sqrt{-g}}
\frac{\partial}{\partial g_{\mu\nu}} ( \sqrt{-g}{ \cal{L}}_{matter}   )$,~
$J^{\nu}=\frac{1}{16\pi b}\frac{\partial} {\partial A_{\mu}}
( { \cal{L}}_{matter}  )$~ and
{}~$\Sigma=\frac{1}{16\pi }\frac{\partial} {\partial \phi}
( { \cal{L}}_{matter} )$ are source terms of gravity, electromagnetic and
dilaton field.

The post-Newtonian limit of the theory is determination of the metric tensor to
order $O(v^4)$ for $g_{00}$, $O(v^3)$ for $g_{0i}$ and $O(v^2)$ for
$g_{ij}$~\cite{37}.  Thus the metric can be written as :
\begin{eqnarray}
g_{00} &=& -1+ {^{(2)}g}_{00}+{^{(4)}g}_{00}+...  ,\nonumber\\
g_{ij} &=& \delta_{ij}+{^{(2)}g}_{ij}+... ,\nonumber\\
 g_{0i}&=& {^{(3)}g}_{0i}+...,
\end{eqnarray}
where $^{(n)}g_{\mu\nu}$ is of order
$v^n$,  and the Latin indices denote the space components which is running
from 1 to 3.
The inverse metric has also an expansion which may be written as
\begin{eqnarray}
g^{00}&=&-1+ {^{(2)}g}^{00}+ {^{(4)}g}^{00}+... ,\nonumber\\
g^{ij}&=&\delta^{ij}+ {^{(2)}g}^{ij}+... ,\nonumber\\
g^{0i}&=& {^{(3)}g}^{0i}+...,
\end{eqnarray}
with
\[ {^{(2)}g}^{00}=- {^{(2)}g}_{00},{^{(2)}g}^{ij}=- {^{(2)}g}_{ij},
{^{(3)}g}^{0i}={^{(3)}g}_{0i} .\]

It is well known that we can choose the harmonic coordinate system which the
metric satisfies the harmonic condition~\cite{37},,
\begin{equation}
 g^{\mu\nu}\Gamma^{k}_{\mu\nu}=0 .
\end{equation}
Under these conditions the Recci tensor were simplified to
\begin{eqnarray}
 {^{(2)}R}_{00} & = & -\frac{1}{2}\nabla^2{^{(2)}g}_{00}, \\
 {^{(2)}R}_{ij} & = & -\frac{1}{2}\nabla^2{^{(2)}g}_{ij}, \\
 {^{(3)}R}_{0i} & =  & -\frac{1}{2}\nabla^2{^{(3)}g}_{0i}, \\
 {^{(4)}R}_{00} & = & \frac{1}{2}{\partial_t}^2{^{(2)}g}_{00}
-\frac{1}{2}\nabla^2{^{(4)}g}_{00}- \frac{1}{2}
\delta^{ij}\partial_i{^{(2)}g}_{00}\partial_j{^{(2)}g}_{00} \nonumber \\
 & & + \frac{1}{2}{^{(2)}g}_{ij}\partial_i\partial_j{^{(2)}g}_{00}.
\end{eqnarray}

Now , we assume the expansion for the dilaton is given by
\begin{equation}
 \phi = {^{(0)}\phi}+ {^{(2)}\phi}+ {^{(4)}\phi}+...   ~~.
\end{equation}
${^{(0)}\phi}$ should be the asymptotical value $\phi_0$ of the dilaton field.
And the expansion for the electromagnetic field is given by
\begin{eqnarray}
 F_{oi} &=&{^{(2)}F}_{oi}+{^{(4)}F}_{oi}... ,\nonumber\\
 F_{ij} &= & {^{(3)}F}_{ij}+...~~.
\end{eqnarray}
It means that, when the Lorentz gauge,
$\partial_\mu A^\mu=0$, was taken, the electric potential $A_0$ and vector
potential $A_i$ have the expansion
\begin{eqnarray}
A_{o} &=& {^{(2)}A}_{o}+{^{(4)}A}_{o}... ,\nonumber\\
A_{i} &=& {^{(3)}A}_{i}+...~~.
\end{eqnarray}
The reason to assign those orders of the field expansion is that we can easy to
control the orders of fields.

We may also expand the energy momentum $T_{\mu\nu}^{m}$, current $J^\mu$ and
dilaton  source $\Sigma$ in powers of $v$
\begin{eqnarray}
T^{00}_m &=& {^{(0)}T}^{00}_m+ {^{(2)}T}^{00}_m+... ,\nonumber\\
T^{0i}_m &=& {^{(1)}T}^{0i}_m+... ,\nonumber\\
T^{ij}_m &=&  {^{(2)}T}^{ij}_m+...,\\
J_{0} &=& {^{(0)}J}_{0}+{^{(2)}J}_{0}... ,\nonumber\\
J_{i} &=& {^{(1)}J}_{i}+...~~, \\
\Sigma &=& {^{(0)}\Sigma}+ {^{(2)}\Sigma}+ ...   ~~.
\end{eqnarray}

Plugging all the expansions into the field equations, and comparing both side
of the equations order by order, we obtain a set of Poisson equations in the
post-Newtonian limit,
\begin{eqnarray}
{\nabla^2}{^{(2)}g}_{00} & = &
-\kappa^2 {^{(0)}T}^{00}_m ,\\ {\nabla^2}{^{(2)}g}_{ij} & = &
-\kappa^2 \delta_{ij}{^{(0)}T}^{00}_m,\\
 {\nabla^2}{^{(2)}g}_{0i} & = &
 {\kappa^2} {^{(1)}T}^{0i}_m, \\
 {\nabla^2}{^{(4)}g}_{00}  & = & -\kappa^2
\bigl( {^{(2)}T}^{00}_m+{^{(2)}T}^{ii}_m-2{^{(2)}g}_{00}{^{(0)}T}^{00}_m  +4b
e^{-2a\phi_0} {^{(2)}F}_{oi}{^{(2)}F}_{oi} \bigr) \nonumber\\
& & +
\partial_0^2{^{(2)}g}_{00}-\partial_i{^{(2)}g}_{00}\partial_i{^{(2)}g}_{00}
+{^{(2)}g}_{ij}\partial_i\partial_j{^{(2)}g}_{00},\\
 {\nabla^2}{^{(2)}\phi} & = & -4\pi{^{(0)}\Sigma},\\
 {\nabla^2}{^{(4)}\phi} & = & -4\pi{^{(2)}\Sigma}+ \partial_0^2{^{(2)}\phi}
\nonumber\\
& & + {^{(2)}g}_{ij}\partial_i\partial_j{^{(2)}\phi} + abe^{-2a\phi_0}
{^{(2)}F}_{oi}{^{(2)}F}_{oi} ,\\
 {\nabla^2}{^{(2)}A}_{0} & = & -4\pi e^{-2a\phi_0} {^{(0)}J}_{0} ,\\
 {\nabla^2}{^{(3)}A}_{i} & = & -4\pi e^{-2a\phi_0} {^{(1)}J}_{i} ,\\
 {\nabla^2}{^{(4)}A}_{0} & = & -4\pi e^{-2a\phi_0} {^{(2)}J}_{0} +
 \partial_0^2 {^{(2)}A}_{0} +  2a{^{(2)}\phi} \nabla^2 {^{(2)}A}_{0}
\nonumber\\
& &- 2a\partial_i {^{(2)}\phi}{^{(2)}F}_{oi} + \partial_i {^{(2)}g}_{00}
{^{(2)}F}_{oi} .
\end{eqnarray}

In order to avoid some obscure energy density terms in $O(v^4)$ equations which
will result in divergences after the integration. The metric, dilaton and
electromagnetic fields are usually rewritten as
\begin{eqnarray}
{^{(2)}g}_{00} & = & -2U, \\
{^{(2)}g}_{ij} & = & -2U \delta_{ij},\\
{^{(2)}\phi} & = & \Theta, \\
{^{(2)}A}_{0} & = & \Phi, \\
{^{(3)}g}_{0i} & = & \xi_{i},\\
{^{(3)}A}_{i} & = & \zeta_{i},\\
{^{(4)}g}_{00} & = & -2\Psi-2U^2-2b \kappa^2
e^{-2a\phi_{0}}\Phi^2~,\\
{^{(4)}\phi} & = & \Xi+\frac{1}{2} ab e^{-2a\phi_{0}}\Phi^2~,\\
{^{(4)}A}_{0} & = &\Upsilon+a\Theta\Phi+U\Phi~.
\end{eqnarray}

After replacing eqs (37)-(45) into eqs. (28)-(36), we end up with a set of
simpler field equations,
\begin{eqnarray}
{\nabla^2}U & = &
  \frac{\kappa^2}{2} {^{(0)}T}^{00}_m ,\\
{\nabla^2}U\delta_{ij} & = &
  \frac{\kappa^2}{2} \delta_{ij}{^{(0)}T}^{00}_m,\\
{\nabla^2}\xi_i & = &
 -{\kappa^2} {^{(1)}T}^{0i}_m, \\
{\nabla^2}\Psi  & = & -\frac{\kappa^2}{2}
 \bigl( {^{(2)}T}^{00}_m+{^{(2)}T}^{ii}_m
    +16\pi b  e^{-4a\phi_0} \Phi {^{(0)}J}_0
    +\frac{2}{\kappa^2}\partial_0^2U \bigr),\\
{\nabla^2}\Theta & = & -4\pi{^{(0)}\Sigma},\\
{\nabla^2}\Xi & = &
     -4\pi ( {^{(2)}\Sigma}- 2U{^{(0)}\Sigma}
     + ab e^{- 2a\phi_0}\Phi{^{(0)}J}_0
     -\frac{1}{4\pi} \partial_0^2\Theta ) ,\\
{\nabla^2}\Phi & = & -4\pi e^{-2a\phi_0} {^{(0)}J}_{0} ,\\
{\nabla^2}\zeta_i & = & -4\pi e^{-2a\phi_0} {^{(1)}J}_{i} ,\\
{\nabla^2}\Upsilon & = & -4\pi \bigl( e^{-2a\phi_0} {^{(2)}J}_{0}
   + a e^{-2a\phi_0}\Theta{^{(0)}J}_{0} -e^{-2a\phi_0} U{^{(0)}J}_{0}
  \nonumber\\& & ~~~~~~+ a \Phi {^{(0)}\Sigma}
- \frac{\kappa^2}{8\pi}\Phi{^{(0)}T_m}^{00}
      -\frac{1}{4\pi}\partial_0^2\Phi \bigr).
\end{eqnarray}

Integrating equations (46)-(54), we find out the solutions for this system :

\begin{eqnarray}
U({\bf x},t)  & = & - {\kappa^2\over 8\pi}\int \frac {d^3x' }
   {|{\bf x}-{\bf x'}|} {^{(0)}T}^{00}_m ({\bf x'},t)  ,\\
\xi_i({\bf x},t) & = & - {\kappa^2\over 4\pi}\int\frac{d^3x'}
     {|{\bf x}-{\bf x'}|} {^{(1)}T}^{0i}_m ({\bf x'},t) ,  \\
\Psi({\bf x},t) & = & - {\kappa^2\over 8\pi}\int \frac{d^3x' }
    {|{\bf x}-{\bf x'}|}
   \bigl[ {^{(2)}T}^{00}_m ({\bf x'},t) + {^{(2)}T}^{ii}_m ({\bf x'},t)
\nonumber\\ & & ~~~~~~~~~~~~~~~~~~~~
  + 16\pi b  e^{-4a\phi_0} \Phi ({\bf x'},t) {^{(0)}J}_0 ({\bf x'},t)
      +{2\over \kappa^2}{\partial_0^2 U} ({\bf x'},t)\bigr],
     \nonumber\\& &\\
\Theta({\bf x},t) & = & \int \frac{d^3x' } {|{\bf x}-{\bf x'}|}
     {^{(0)}\Sigma}({\bf x'},t) ,\\
\Xi ({\bf x},t) & = & \int \frac{d^3x' } {|{\bf x}-{\bf x'}|}
      \bigl[ {^{(2)}\Sigma}({\bf x'},t)
        - 2U({\bf x'},t){^{(0)}}\Sigma({\bf x'},t)
  \nonumber\\ & & ~~~~~~~~~~~~~~~
     + ab e^{- 2a\phi_0}\Phi ({\bf x'},t){^{(0)}J}_0 ({\bf x'},t)
     -\frac{1}{4\pi} \partial_0^2\Theta({\bf x'},t) \bigr], \\
\Phi({\bf x},t) & = & \int \frac{d^3x' }
     {|{\bf x}-{\bf x'}|} e^{2a\phi_0} {^{(0)}J}_{0}({\bf x'},t) ,\\
\zeta_i({\bf x},t) & = & \int \frac{d^3x' }
    {|{\bf x}-{\bf x'}|} e^{2a\phi_0} {^{(1)}J}_{i}({\bf x'},t) ,\\
\Upsilon({\bf x},t) & = & \int \frac{d^3x' }
   {|{\bf x}-{\bf x'}|}  \bigl[ e^{-2a\phi_0} {^{(2)}J}_{0}({\bf x'},t)
   + a e^{-2a\phi_0}\Theta({\bf x'},t){^{(0)}J}_{0}({\bf x'},t)
   \nonumber\\ & & ~~~~~~~~~~~~~~~~
   -e^{-2a\phi_0} U({\bf x'},t){^{(0)}J}_{0}({\bf x'},t)
   + a \Phi({\bf x'},t) {^{(0)}\Sigma}({\bf x'},t)
      \nonumber\\ & & ~~~~~~~~~~~~~~~~
    - \frac{\kappa^2}{8\pi}\Phi({\bf x'},t){^{(0)}T_m}^{00}({\bf x'},t)
      -\frac{1}{4\pi}\partial_0^2\Phi ({\bf x'},t) \bigr]~~.
\end{eqnarray}

Here, when the source distributions are given, all the post-Newtonian expansion
of metric, dilaton and electric fields can be carried out by straightforward
integrations.

Let us illustrate this scheme in a simple example, the approximate solution for
a point mass ${\cal M}$ which carries electric $Q$ and dilaton charges $D$. For
this case, the source terms are :
\[ {^{(0)}T}^{00}_m={\cal M}{\delta^3({\bf x})},\]
\[ {^{(0)}\Sigma}=-D{\delta^3({\bf x})},\]
\[ {^{(0)}J}_{0}=Q{\delta^3({\bf x})},\]
\begin{equation}
{^{(2)}T}^{00}_m= {^{(1)}T}^{0i}_m={^{(2)}T}^{ij}_m=0~~, {^{(2)}J}_{0}=
{^{(1)}J}_{i}=0~~,{^{(2)}\Sigma}=0  ~~.
\end{equation}

Putting into the integral, finally, we get the results :
\begin{eqnarray}
U({\bf x}) & = & - \frac{ M} {|{\bf x}|},\nonumber\\
\xi_i({\bf x}) & =& 0,\nonumber\\
\Psi({\bf x}) & = & 0,\nonumber\\
\Theta({\bf x}) & = & -\frac{D } {|{\bf x}|},\nonumber\\
\Xi({\bf x}) & = & 0 ,\nonumber\\
\Phi({\bf x}) & = & \frac{Q e^{2a\phi_0} } {|{\bf x}|},\nonumber\\
\zeta_i({\bf x}) & = & 0 ,\nonumber\\
\Upsilon({\bf x}) & = &0 ,
\end{eqnarray}
where $M={\kappa^2 \over 8\pi}{\cal M}$.

Therefore the post-Newtonian expansion of metric, dilaton and  electric field
strengths are
\begin{eqnarray}
g_{00} &=& -\{ 1-\frac{2M}{|{\bf x}|}
+\frac{1}{|{\bf x}|^2}(2M^2+{2\kappa^2 b Q^2 e^{2\phi_0} })+...\}~~.\nonumber\\
g_{ij} &=& \delta_{ij}+\frac{2M}{|{\bf x}|}\delta_{ij}
,\nonumber\\
g_{0i} &=& 0.\\
F_{0i} &=& \frac{x_i}{|{\bf x}|^3}{Q e^{2\phi_0} }
   -\frac{x_i}{|{\bf x}|^4}{Q e^{2a\phi_0} } (2M+2aD)
    +...~~,\\
{\phi} &=&\phi_0 -\frac{D } {|{\bf x}|}+{ab \over 2}
    \frac{Q^2 e^{2\phi_0} } {|{\bf x}|^2} +... ~~.
\end{eqnarray}

Besides the asymptotical background,  the lowest order of potential expansions,
${^{(2)}g}_{00}, {^{(2)}A}_{0}$ and ${^{(2)}\phi}$, are order of $v^2$. It
means that all the charges $Q$, $D$ and mass $M$ are same order as
$\phi_0\approx 0$ in this example.

\vspace{10mm}
\section{An electric dilatonic black hole in the harmonic coordinate}

    In order to make sure those results presented in the last section are
correct.  We will compare the approximate solution of a point mass which
carries electric and dilaton charges to the pure, electric dilatonic black hole
solution of dilaton gravity.

     From the exact solution, eqs (2)-(9),  we know that the electric dilatonic
black hole is also characterized by a dilaton charge~\cite{14}
\begin{equation}
D={1\over 4\pi}\oint_{S} d^2S^\mu\nabla_\mu\phi
=\frac {Q^2 e^{2a\phi_0} } {2M}.
\end{equation}
Due to the dilaton source is free in the dilaton gravity, eq(1), $D$ is not a
new free parameter in this case.  Once $\phi_0$ is fixed, it is determined by
$M$ and $Q$. Here we have set $Q=Q_e$ .

For convenience, we rewrite the charged black hole solution, eq.(2), in terms
of dilaton charge, electric charge and mass, as followings:
\begin{equation}
   ds^2 = -\Delta^2 dt^2 + \frac{\sigma^2}{\Delta^2} dr^2
          + r^2 \left( {d\theta}^2
          + {\sin}^2 \theta {d\varphi}^2 \right)~~,
\end{equation}
where
\begin{equation}
   \Delta^2 = 1 - \frac{2 M}{r^2} \sqrt{r^2 + D^2}
                + \frac{{Q^2 e^{2\phi_0} }}{r^2}~~,
\end{equation}
\begin{equation}
   \sigma^2 = \frac{r^2}{r^2 + D^2}~~,
\end{equation}
 The electric field is
\begin{equation}
   F_{01} = \frac{Q}{r^2} e^{2\phi}~~,
\end{equation}
and the dilaton field obeys
\begin{equation}
e^{2\phi} = e^{2\phi_0}(1-\frac{2 D}{\sqrt{r^2 + D^2} + D})~~.
\end{equation}

Moreover, we change the standard spherical coordinates $(t, r, \theta,
\varphi)$ to harmonic coordinates $(t, x_i)$, which satisfy the harmonic
condition~\cite{37},
\begin{equation}
g^{\mu\nu}\nabla_\mu\nabla_\nu x_i=0~~.
\end{equation}
If we choose the transformations between the harmonic coordinates  and the
standard coordinates to be
\begin{eqnarray}
x_1 &=& R(r)sin\theta cos\varphi,\nonumber\\
x_2 &=& R(r)sin\theta sin\varphi,\nonumber\\
x_3 &=& R(r) cos\varphi,
\end{eqnarray}
then the transformation function $R(r)=|{\bf x}|$ is determined by
\begin{equation}
\frac{d}{dr} (r^2 {\Delta^2\over \sigma}{dR\over dr})
-2\sigma R=0~~.
\end{equation}
In the far region, the approximate solution of $R(r)$ is
\begin{equation}
R(r)=r(1-{M\over r}+ \frac {1}{2}{D^2\over r^2}+...)~~.
\end{equation}

After some complicate calculation,  we get the weak field expansions of the
metric in the harmonic coordinate,
\begin{eqnarray}
g_{00} &=& -\{ 1-\frac{2M}{|{\bf x}|}
+\frac{1}{|{\bf x}|^2}(2M^2+{Q^2 e^{2\phi_0} })+...\}~~.\nonumber\\
g_{ij} &=& \delta_{ij}+\frac{2M}{|{\bf x}|}\delta_{ij}
,\nonumber\\
g_{0i} &=& 0.
\end{eqnarray}
 The electric and dilaton field expansion are
\begin{eqnarray}
F_{0i} &=& \frac{x_i}{|{\bf x}|^3}{Q e^{2\phi_0} }
   -2\frac{x_i}{|{\bf x}|^4}{Q e^{2a\phi_0} } (M+D) +...~~,\\
{\phi} &=&\phi_0 -\frac{D } {|{\bf x}|}+{1 \over 2} \frac{Q^2 e^{2\phi_0} }
{|{\bf x}|^2} +... ~~.
\end{eqnarray}
Obviously, these expansions are consistent with the approximate solution of
dilaton gravity, ($2\kappa^2=a=b=1$), which was found in the post-Newtonian
scheme.

\vspace{10mm}
\section{Concluding remarks}

we derived the field equation of the generalized dilaton gravity in the
post-Newtonian limit . We also demonstrated an approximate solution for a point
mass with electric charge and dilaton charge, and compare it with the weak
field expansion of a charged, dilatonic black hole in the harmonic coordinate.

According to eqs(43),(57) or (65), the dilaton charge does not give any
contribution to metric up to post-Newtonian terms, $^{(4)}g_{00}$.   Therefore,
a neutral test particle, which obey the equation of motion
\begin{equation}
\frac{d^2x^\mu}{d\tau^2}+\Gamma^\mu_{\alpha\beta}\frac{dx^\alpha}{d\tau}\frac{dx^\beta}{d\tau}=0,
\end{equation}
can not detect the dilaton effect in the post-Newtonian level.
But, the dilaton charge gives post-Newtonian contribution to the electric
field, $^{(4)}F_{0i}$, see eqs (45),(60),(62) or (66).  Whereas a charged test
particle, which obey the equation of motion
\begin{equation}
\frac{d^2x^\mu}{d\tau^2}+\Gamma^\mu_{\alpha\beta}\frac{dx^\alpha}{d\tau}\frac{dx^\beta}{d\tau}=\frac{e}{m}F^\mu_\nu\frac{dx^\nu}{d\tau},
\end{equation}
can detect the effects of the dilaton charge.

However, a charged particles test is hard to be assembled~\cite{34}\cite{36},
we still need to calculate the post-post-Newtonian terms for the neutral
particle test.  The post-post-Newtonian limit is under our current
investigation.  We are also interesting in studying the post-Newtonian
approximate solutions for the more general dilaton gravity in which a mass term
$m\phi^2$ or potential term $V(\phi)$  are included.

\vspace{20mm}
\begin{center}
{\large \bf Acknowledgements}
\end{center}
\vspace{10mm}
We thank Prof. W.-T. Ni, Prof C.-C. Chen and Prof.T.-H. Cho for usefull
discussion. This work was supported in part by the National Science Council of
the Republic of China under grants No.~NSC 84-2112-M006-010,  No.~NSC
85-2112-M006-005 and No.~NSC 84-2112-M194-003.
\newpage

\end{document}